\documentclass[amsfonts, prd, twocolumn, nofotinbib, showpacs]{revtex4}

\usepackage{amssymb}
\usepackage{amsmath}
\usepackage{amsfonts}
\usepackage{graphicx, float}
\usepackage{graphicx, epsfig}
\usepackage{color}
\usepackage{enumerate}
\usepackage{latexsym}
\usepackage{graphicx, epsfig}
\usepackage{color}
\usepackage{enumerate}
\newcommand{\be}{\begin{equation}}
\newcommand{\ee}{\end{equation}}
\newcommand{\beq}{\begin{equation}}
\newcommand{\eeq}{\end{equation}}
\newcommand{\bea}{\begin{eqnarray}}
\newcommand{\eea}{\end{eqnarray}}

\newcommand{\Mbh}{M_{\rm BH}}

\begin{document}

\title{Black Hole Thermodynamics in MOdified Gravity (MOG)}

\author{Jonas R. Mureika$^1$, John W. Moffat$^{2,3}$, Mir Faizal$^3$}
\affiliation{$^1$ Department of Physics, Loyola Marymount University, Los Angeles, CA, 90045~~USA\\
$^2$ Perimeter Institute for Theoretical Physics, Waterloo, Ontario,  N2L 2Y5~~Canada\\
$^3$ Department of Physics and Astronomy, University of Waterloo, Ontario, N2L 3G1~Canada}

\begin{abstract}
We analyze the thermodynamical properties of black holes in a modified theory
of gravity, which was initially proposed to obtain correct dynamics
of galaxies and galaxy clusters without dark matter.
The thermodynamics of non-rotating and rotating black hole
solutions resembles similar solutions in
Einstein-Maxwell theory with  the electric charge being replaced by
a new mass dependent gravitational charge $Q = \sqrt{\alpha G_N}M$.
This new mass dependent charge modifies the effective Newtonian constant from $G_N$ to $G = G_N(1+\alpha)$,
and this in turn critically affects the thermodynamics  of the black holes.
We also investigate the thermodynamics of regular solutions, and explore the limiting case when no horizons forms.
So, it is possible that the modified gravity can lead to the absence of
black hole  horizons in our universe.  Finally, we analyze corrections to the thermodynamics
of a non-rotating black hole and obtain the usual logarithmic correction term. 
\end{abstract}

\maketitle

\section{Introduction}

General relativity is one of the most successful and well-tested theories of gravitation to date.  Nevertheless, there is still 
motivation to modify the framework at large scales. One of the main reasons is the discrepancy between the observed dynamics of galaxies and clusters of galaxies and the amount of luminous matter these galaxies and clusters contain~\cite{rubin1,rubin2,Zwicky}. This is usually
explained by postulating the existence of exotic dark matter. To date, no dark matter particle candidates have been
detected in laboratory experiments or in satellite missions. An alternative resolution to the problem of the
galaxy and galaxy cluster dynamics is a modification of the laws of gravitation on scales where Newtonian
gravity or general relativity (GR) have not been extensively tested. One such framework called MOG
(MOdified Gravity)~\cite{Moffat1} has been able to explain the dynamics of galaxies and galaxy clusters without 
the need for dark matter in the present epoch of the universe~\cite{MoffatRahvar1,MoffatRahvar2,BrownsteinMoffat1,MoffatToth}.

In one version of the MOG formulation called the Scalar-Tensor-Vector Gravity (STVG) theory~\cite{Moffat1},
the field content of general relativity has been increased to include scalar fields and a massive vector field.
It has been used to describe the growth of structure, the matter power spectrum and the cosmic microwave background
(CMB) acoustical power spectrum data in the early universe~\cite{Moffat2,Moffat3}. Solar system experiments are also
in accordance with MOG~\cite{Moffat4}, and the dynamics of the Bullet Cluster has been explained without dark
matter~\cite{BrownsteinMoffat2}. It may be noted that even though MOG fits the galaxy rotation curves like
the phenomenological MOND~\cite{mond1}, the advantage of using MOG is that it is constructed from a fully
covariant modification of the Einstein-Hilbert action in GR. Thus, it is possible to apply MOG to cases
where the gravitational field is strong, and the weak field approximation breaks down. It may be noted
that there are various other ways to obtain the modified Newtonian potential from a covariant formalism. In fact, one such approach proposed by 
Bekenstein, called Tensor-Vector-Scalar (TeVeS) gravity produces MOND in the weak field
approximation~\cite{Bekenstein}.

In this paper, we will analyze black hole thermodynamics in MOG. It may be noted that an entropy has to be associated with 
black holes in order to prevent the violation of the second law of thermodynamics \cite{1, 1a}. 
Black holes also have a temperature and they evaporate due to Hawking radiation. 
However, it has been demonstrated that the  evaporation of asymptotically de Sitter 
black holes (whose size is comparable to that of the cosmological horizon)  differs 
 significantly from the evaporation of asymptotically flat black holes \cite{bh12}.  In fact, it has been observed 
 that it is possible for quantum Schwarzschild-de Sitter black holes to  anti-evaporate. 
 The evaporation and anti-evaporation of black holes has also been analysed in $f(T)$ gravity \cite{bh14}
 and $F(R)$ gravity \cite{bh15}. The black hole entropy gets generalized to Wald entropy \cite{apqw1, apqw2, apqw4, apqw5}. 
 The entropy formula can be used to compute the entropy of a black hole in any 
modified theory of gravity.  The black hole entropy has been studied in Gauss-Bonnet gravity \cite{gauss1, gauss2}, 
and Lovelock gravity \cite{lock1, lock2}.  The black hole thermodynamics has also been studied in massive gravity \cite{massive, massive2}. 

The black holes have more entropy than any other object of the same volume~\cite{2, 4}, and this 
 maximum  entropy scales with the area of the black hole~\cite{4a}. This observation
has led to the development of the holographic principle~\cite{5, 5a}. 
However, as  the black holes get smaller in size due to Hawking radiation, quantum fluctuations are expected to 
correct the standard relation between the area and the entropy of a black hole.
This in turn can modify the holographic principle~\cite{6, 6a}.
Such corrections have been evaluated using several different approaches. Non-perturbative quantum general
relativity has been used to calculate the density of microstates for asymptotically
flat black holes~\cite{1z}. Here, the density of states has been obtained
by using a suitable conformal field theory. It has been demonstrated
using the Cardy formula that all black holes whose microscopic degrees of freedom are described by a
conformal field theory will have logarithmic corrections~\cite{card,ca}. The exact partition function has been computed for
BTZ black holes, and the logarithmic corrections have been obtained from an exact
partition function~\cite{card}. Such correction have also been obtained using matter fields in
backgrounds of a black hole~\cite{other,other0,other1}. The logarithmic corrections are also produced from
string theoretical effects~\cite{solo1,solo2,solo4,solo5,jy}. Such corrections can also be generated from 
thermal fluctuations that occur in the thermodynamics of black holes \cite{l1, SPR}. We will analyse the corrections due to thermal 
fluctuations for a MOG black hole, and observe that a MOG black hole also receives logarithmic corrections.  

In this paper, we will focus on static and rotating black hole solutions in MOG~\cite{Moffat5,Moffat6}, and will analyze their thermodynamics 
in the standard fashion.  We show that the Hawking temperature and evaporation profile of the static MOG black hole is significantly modified 
over the standard Schwarzschild case.  The rotating black hole is shown to admit remnant behavior. 
We also present corrections to our derived thermodynamic expressions, and find the usual logarithmic form.

\section{Modified Gravity Action and Field Equations}

The action in the STVG formulation of MOG can be written as~\cite{Moffat1}:
\begin{equation}
\label{action1}
S=S_G+S_\phi+S_S+S_M.
\end{equation}
Here, $S_G$ is the original Einstein-Hilbert gravity action, $S_\phi$ is the action of a massive vector field
$\phi_\mu$, $S_S$ is the action of scalar fields and $S_M$ is the action for pressure-less matter:
\begin{eqnarray}
S_G&=&\frac{1}{16\pi}\int\frac{1}{G}\left({\it R}+2\Lambda\right)\sqrt{-g}~d^4x, \\
 S_\phi&=&-\frac{1}{4\pi}\int\Big[\mathcal{K} + V(\phi_\mu)\Big]\sqrt{-g}~d^4x,  \\
S_S&=&\int\frac{1}{G}\Big[\frac{1}{2}g^{\alpha\beta}\biggl(\frac{\nabla_\alpha G\nabla_\beta G}{G^2}
+\frac{\nabla_\alpha\mu\nabla_\beta\mu}{\mu^2}\biggr)\nonumber\\
 && -\frac{V_G(G)}{G^2}-\frac{V_\mu(\mu)}{\mu^2}\Big]\sqrt{-g}~d^4x,  \\
S_M &=& -\int(\rho \sqrt{u^\mu u_\mu}+Qu^\mu\phi_\mu)\sqrt{-g}~d^4x +J^\mu\phi_\mu,
\end{eqnarray}
where $R=g^{\mu\nu}R_{\mu\nu}$, $g={\rm det} (g_{\mu\nu})$, $\nabla_\mu$ is the covariant derivative with respect to the metric $g_{\mu\nu}$,
the $Vs$ are potential contributions to the action and we use units for which $c=1$ and the metric signature is $(+,-,-,-)$.
Moreover, $ \mathcal{K}$ is the kinetic term for the $\phi_\mu$ field. It is possible to 
choose it to be the usual kinetic term for a vector field, 
\begin{equation}
\mathcal{K}=\frac{1}{4}{\bf\it B^{\mu\nu}B_{\mu\nu}},
\end{equation}
where
\begin{equation}
B_{\mu\nu}=  \partial_\mu \phi_\nu - \partial_\nu \phi_\mu.
\end{equation}

The STVG field equations are given by~\cite{Moffat1}:
\begin{equation} 
\label{mog1}
G_{\mu\nu}-\Lambda g_{\mu\nu}+Q_{\mu\nu}=-8\pi GT_{\mu\nu},
\end{equation}
where
\begin{equation}
Q_{\mu\nu}=G(\Box\Theta g_{\mu\nu}-\nabla_\mu\nabla_\nu\Theta),
\end{equation}
where $G_{\mu\nu}=R_{\mu\nu}-1/2g_{\mu\nu}R$, $\Lambda$ is the cosmological constant and $\Theta=1/G$.

The field equations for $B^{\mu\nu}$ are given by
\begin{equation}
\nabla_\nu B^{\mu\nu}+\frac{\partial V(\phi_\mu)}{\partial\phi_\mu}=-J^\mu,
\end{equation}
and
\begin{equation}
\nabla_\sigma B_{\mu\nu}+\nabla_\mu B_{\nu\sigma}+\nabla_\nu B_{\mu\sigma}=0.
\end{equation}
Moreover, we have the field equations
\begin{equation}
\label{mog4}
\Box G=K(x),
\end{equation}
\begin{equation}
\label{mog5}
\Box\mu=L(x),
\end{equation}
where $\Box=\nabla^\alpha\nabla_\alpha$ and
\begin{equation}
K(x)=\biggl(\frac{16\pi}{3+16\pi}\biggr)\biggl[\frac{3}{8\pi G}(1+4\pi)\nabla^\alpha G\nabla_\alpha G 
$$ $$
- \frac{G}{2\mu^2}\Box\mu+\frac{1}{2}G^2\biggl(T+\frac{\Lambda}{4\pi G}\biggr)+\frac{1}{\sqrt{\alpha G_N}}T^{M\mu\nu}u_\nu\phi_\mu\biggr],
\end{equation} 
and
\begin{equation}
L(x)=\frac{1}{G}\nabla^{\alpha}G\nabla_{\alpha}\mu+\frac{2}{\mu}\nabla^\alpha\mu\nabla_\alpha\mu+\mu^2G\frac{\partial V(\phi_\mu)}{\partial \mu}.
\end{equation}

The covariant current density is defined to be
\begin{equation}
\label{currentdensity}
J^\mu=\kappa T^{M\mu\nu}u_\nu,
\end{equation}
where $T^{M\mu\nu}$ is the energy-momentum tensor for matter and $\kappa=\sqrt{\alpha G_N}$, $\alpha=(G-G_N)/G_N$ is a scalar field, $G_N$ is Newton's constant, $u^\mu=dx^\mu/ds$ and $s$ is the proper time along a particle trajectory.
The perfect fluid energy-momentum tensor for matter is given by
\begin{equation}
\label{energymomentum}
T^{M\mu\nu}=(\rho_M+p_M)u^\mu u^\nu-p_Mg^{\mu\nu},
\end{equation}
where $\rho_M$ and $p_M$ are the density and pressure of matter, respectively.  We get from (\ref{currentdensity}) and (\ref{energymomentum}) by using $u^\nu u_\nu=1$:
\begin{equation}
J^\mu=\kappa\rho_M u^\mu.
\end{equation}
The gravitational source charge is given by
\begin{equation}
Q=\kappa\int d^3x J^0(x).
\end{equation}

The values $Q=\sqrt{\alpha G_N}M$ and $G=G_N(1+\alpha)$ originate in the weak field approximation of the STVG field equations. The weak field approximation is based on a perturbation about the Minkowski metric $\eta_{\mu\nu}$:
\begin{equation}
g_{\mu\nu}=\eta_{\mu\nu}+\lambda h_{\mu\nu}.
\end{equation}
The test particle equation of motion is given by
\begin{equation}
\frac{d^2x^\mu}{ds^2}+{\Gamma^\mu}_{\alpha\beta}\frac{dx^\alpha}{ds}\frac{dx^\beta}{ds}=\frac{q}{m}{B^\mu}_\nu\frac{dx^\nu}{ds},
\end{equation}
where $m$ and $q=\sqrt{\alpha G_N}m$ are the test particle mass and gravitational charge, respectively, and $\phi_\mu=(\phi_0,\phi_i)\,(i=1,2,3)$.  Assuming that $V(\phi_\mu)$ is given by  
\begin{equation}
V(\phi_\mu)=-\frac{1}{2}\mu^2\phi^\mu\phi_\mu,
\end{equation}
and $\partial_\nu\phi^\nu=0$, the weak field spherically symmetric static, source-free solution for $\phi_0(r)$ is obtained from the equation ($\phi_0'=d\phi_0/dr$):
\begin{equation}
\phi_0''+\frac{2}{r}\phi_0'-\mu^2\phi_0=0.
\end{equation}
The solution is given by
\begin{equation}
\label{phisolution}
\phi_0(r)=-Q\frac{\exp(-\mu r)}{r},
\end{equation}
where the gravitational charge $Q=\sqrt{\alpha G_N}M$ and $M$ is the mass of the source particle.

In the slow motion and weak field approximation, $dr/ds\sim dr/dt$ and $2GM/r\ll 1$, and for the radial acceleration of the test particle we get
\begin{equation}
\frac{d^2r}{dt^2}+\frac{GM}{r^2}=\frac{qQ}{m}\frac{\exp(-\mu r)}{r^2}(1+\mu r).
\end{equation}
For $ qQ/m=\alpha G_NM$ and $G=G_N(1+\alpha)$, the modified Newtonian acceleration law for a point particle is given by~\cite{Moffat1}:
\begin{equation}
\label{accelerationlaw}
a(r)=-\frac{G_NM}{r^2}[1+\alpha-\alpha\exp(-\mu r)(1+\mu r)].
\end{equation} 
We observe that the acceleration of a particle is independent of its material content (weak equivalence principle). We can rewrite this modified acceleration equation as 
\begin{equation}
\label{effectiveaccel}
a(r)=-\frac{{\cal G}(r)M}{r^2},
\end{equation}
where the effective gravitational coupling strength is given by
\begin{equation}
\label{effectiveG}
{\cal G}(r)=G_N[1+\alpha-\alpha\exp(-\mu r)(1+\mu r)].
\end{equation}

In our generalized gravitational theory electromagnetic waves (photons) and gravitational waves (gravitons) move with the speed of light~\cite{Moffat4}.
The range of values of the parameter $\alpha$ for {\it weak gravitational fields} is locally dependent on the size scale of the astrophysical objects being investigated. A photon path is determined by a null geodesic and the weak field solar system bending of light and the Shapiro time delay experimental data are in agreement with GR for the parameter $\alpha$ having the bound $\alpha\ll 1$~\cite{Moffat1}. The perihelion advance of the planet Mercury's orbit is in agreement with GR for $\alpha\ll 1$~\cite{Moffat1}.
The enhanced gravitational interaction experienced by photons at the scale of galaxies and galactic clusters leads to an explanation of gravitational lensing data without dark matter for $\alpha > 1$.  

In the following, we will use the matter-free MOG field equations~\cite{Moffat5,Moffat6}:
\begin{equation}
\label{MOGgraveqs}
G_{\mu\nu}=-8\pi G T_{\phi\mu\nu},
\end{equation}
\begin{equation}
\nabla_\nu B^{\mu\nu}=0,
\end{equation}
\begin{equation}
\nabla_\sigma B_{\mu\nu}+\nabla_\mu B_{\nu\sigma}+\nabla_\nu B_{\mu\sigma}=0.
\end{equation}
The energy-momentum tensor of matter $T_{M\mu\nu}$ in the gravitational field equations has been set equal to zero and we have
\begin{equation}
\label{Tphi}
T_{\phi\mu\nu}=-\frac{1}{4\pi}({B_\mu}^\alpha B_{\nu\alpha}-\frac{1}{4}g_{\mu\nu}B^{\alpha\beta}B_{\alpha\beta}),
\end{equation}
We have set the cosmological constant $\Lambda$ to zero and neglected the vector field $\phi_\mu$ mass $\mu$. The best fits to galaxy and cluster dynamics yielded $\mu=0.042\,{\rm kpc}^{-1}$ corresponding to a mass $m_\phi=2.6\times 10^{-28}$ eV~\cite{MoffatRahvar1,MoffatRahvar2}. This tiny mass can be neglected for the black holes solutions considered in the following. 

There is no reason to choose the usual kinetic term for the vector field $\phi_\mu$; it is possible to consider a non-linear kinetic term. Even though such a kinetic term will reduce to the usual kinetic term in the low energy limit, it can affect the behavior of black hole solutions. It is possible 
to consider general non-linear kinetic $\mathcal{K}$ terms~\cite{Garcia, non, non1, non2, non4,
non5, non6}:
\begin{eqnarray}
\mathcal{K} = f( \mathcal{B}),
\end{eqnarray}
where
\begin{eqnarray}
 \mathcal{B}=B^{\mu\nu}B_{\mu\nu}. 
\end{eqnarray}
A regular black hole soltion has been constructed by coupling the Einstein action to a non-linear electromagnetic field 
\cite{Garcia}. In fact, such a non-linear kinetic term has also been used in MOG for constructing regular black hole solutions \cite{Moffat5}. 
We will explicitly consider such a non-linear kinetic term in MOG,  without a singularity at radial coordinate $r=0$.
By using a Legendre transformation, we can write  $\mathcal{K} = 
2 P H_p - \mathcal{H}$, where $d \mathcal{H} = H_p dP$. Now for a regular black hole, we can write  
 $\mathcal{H} $ as 
\begin{eqnarray}
\mathcal{H}&=& P\,\frac{\left( 1-3\sqrt{-2\,Q^2P}\right) }{\left( 1+\sqrt{%
-2\,Q^2P}\right) ^3}\nonumber \\ && -\frac 3{2\,Q^2s}\left( \frac{\sqrt{-2\,Q^2P}}{1+\sqrt{%
-2\,Q^2P}}\right) ^{5/2},   
\end{eqnarray}
where $  s = Q/2M = \sqrt{\alpha G_N}/2$,  and $P $ is a negative invariant quantity.  The non-linear kinetic term corresponding to this can be written~\cite{Garcia}:
\begin{eqnarray}
\mathcal{K} &=&P\,\frac{\left( 1-8 \sqrt{-2\,Q^2 P} -6\,Q^2P\right) }{\left( 1+\sqrt{%
-2\,Q^2P}\right) ^4}\nonumber \\ && -\frac 3{4\,Q^2s}\frac{(-2\,Q^2P)^{5/4}\left( 3-2\sqrt{%
-2\,Q^2P}\right) }{\left( 1+\sqrt{-2\,Q^2P}\right) ^{7/2}}.  
\end{eqnarray}
By using these non-linear kinetic terms, we can derive a regular solution~\cite{Moffat5},
without a singularity at radial coordinate $r=0$. Such regular solutions in MOG based on nonlinear dynamics for the $B_{\mu\nu}$ field can possess
two horizons or no horizon, depending on the value of $\alpha$~\cite{Moffat5}. For $\alpha <
\alpha_{\rm crit}=0.673$ a particular regular solution possesses two horizons, while for $\alpha > \alpha_{\rm crit}$
the solution does not possess horizons and is also regular at $r=0$. The latter solution is called a ``grey hole''.
It is expected that Hawking's information loss paradox~\cite{Hawking2,Giddings1,Giddings2} will not occur for
these gray holes. %It is interesting to note that a modification of gravity that was initially proposed to explain
%the dynamics of galaxies and galaxy clusters without dark matter, might predict the absence of horizons 
%for massive collapsed astrophysical bodies in our universe, and thereby solve the black hole information loss paradox.

An electrically neutral black hole in MOG will appear similar to a GR static, spherically symmetric electrically charged Reissner-Nordstr\"om black hole~\cite{Moffat5,Moffat6}, but the MOG black hole solution depends only on the mass $M$ and $\alpha$. The rotating black hole solution in MOG will be similar to the Kerr-Newman solution in GR, but will depend on the mass $M$, the angular momentum $J$ and not the electric charge. The Newtonian constant is modified by $G=G_N(1+\alpha)$, where $\alpha$ is a free parameter. For {\it weak} gravitational fields the value of $\alpha$ is fixed by the fits of the modified Newtonian acceleration law to galaxy rotation curves and galaxy clusters: $\alpha=8.89$~\cite{Moffat2,MoffatRahvar1,
MoffatRahvar2}. However, for the strong gravitational fields of MOG black holes, the weak field determination of $\alpha$ is no longer valid. For a given action for the vector field $\phi_\mu$ and the field strength $B_{\mu\nu}$, the uniqueness theorems for the MOG solutions still hold. 

Real astrophysical bodies including black holes are not electrically charged.  If they do have a small electric charge $Q_e$, it will 
have a negligible effect on the curvature of spacetime~\cite{Gibbons}. For this reason the Reissner-Nordstr\"om black hole is not a physically interesting solution.
On the other hand, the vector field $\phi_\mu$ in MOG is expected to be a physically significant field with positive gravitational charge 
$Q=\kappa M=\sqrt{\alpha G_N}M$. Because there is no negative gravitational charge $Q$, MOG black holes are not 
gravitationally charge neutral. Furthermore, because the gravitational charge in MOG is not independent of the mass of the black hole, 
it is not possible to construct static spherically symmetric extremal solutions where a remnant is left after Hawking radiation 
evaporation~\cite{2,Hawking2}. In fact, in this paper it will be demonstrated that for static spherically symmetric MOG black holes, it is not 
possible to have a solution where the black hole temperature vanishes.

\section{Thermodynamics of Static Non-rotating Black Holes in MOG}

The first solution we will analyze is a static, spherically symmetric
solution to the field equations. The metric for this solution can be written as~\cite{Moffat5,Moffat6}:
\beq
\label{MOG metric}
ds^2 =\left(1-\frac{2GM}{r} + \frac{GQ^2}{r^2}\right) dt^2
$$ $$
- \biggl(1-\frac{2GM}{r} + \frac{GQ^2}{r^2}\biggr)^{-1}dr^2 - r^2 d\Omega^2,
\eeq
where the usual Newtonian constant $G_N$ gets modified to $G = G_N(1+\alpha)$ and $Q=\sqrt{\alpha G_N}M$.
The metric exhibits two horizons:
\beq
r_\pm = G_N M \left(1+\alpha \pm \sqrt{1+\alpha}\right).
\label{moghorizons}
\eeq
We observe that for $\alpha > 0$, in contrast to the Reissner-Nordstr\"om solution when $M < |Q_e|$ ($Q_e$ is the electric charge),
the MOG black hole never has a naked singularity. Thus, as in the case of the Schwarzschild solution,
the MOG black hole satisfies Penrose's cosmic censorship postulate~\cite{Penrose1,Penrose2}.

The temperature of the black hole can be calculated via the surface gravity method evaluated at the outer horizon:
\beq
T = \frac{\kappa_g}{2\pi}~~~,~~~\kappa_g = \frac{1}{2} \frac{d g_{00}}{dr}\left(r=r_+\right),
\label{tsurfgrav}
\eeq
which gives
\beq
T = \frac{1}{2\pi G_NM} \cdot \frac{1}{(1+\sqrt{1+\alpha})(1+\alpha+\sqrt{1+\alpha})}.
\label{mogtemp1}
\eeq
This temperature can also be obtained by using the standard expression
\beq
T = \frac{r_+ - r_-}{4\pi r_+^2}.
\eeq

When $\alpha = 0$, the usual Schwarzschild black hole temperature is obtained, $T = 1/8\pi G_N M$.
It is known that it is possible to construct an extremal solution for the electrically
charged Reissner-Nordstr\"om black hole. When the temperature vanishes $T = 0$ a remnant
exists after the black hole has evaporated by Hawking radiation. Thus, the information loss
paradox can be resolved for these extremal Reissner-Nordstr\"om black hole solutions.
However, as we have explained, because electrically charged black holes are not expected to
be physically reasonable, this is not a viable resolution of the information loss paradox.
However, as we will demonstrate, since an extremal solution does not exist for a static spherically
symmetric MOG black hole as long as $\alpha > 0$, there is no solution for which $T=0$ implying that
there are no remnants remaining after Hawking evaporation.

\begin{figure}[h]
\begin{center}
\includegraphics[scale=0.35]{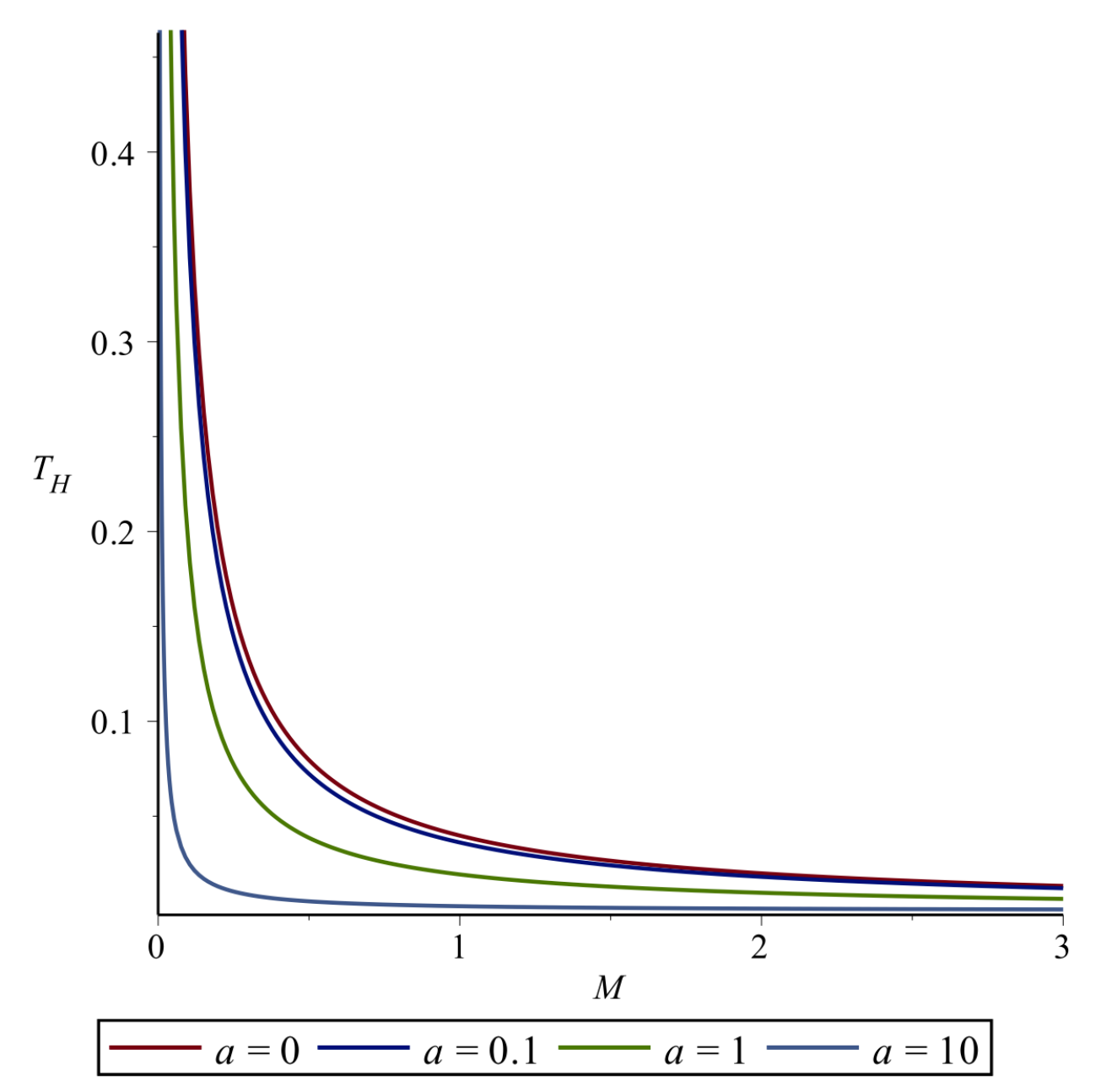}
\caption{Non-rotating MOG black hole Hawking temperature for $\alpha = 10,1,0.1$ (from left to right).
The gravitational constant has been set to $G_N = 1$.  When $\alpha=0$, the temperature
reduces to the Schwarzschild black hole temperature.  Increasing values of the parameter $\alpha$
serve to suppress the radiative power for larger values of $M$, but with increasing power for $M\rightarrow 0$.}
\label{fig1}
\end{center}
\end{figure}

The temperature is plotted in Fig.~\ref{fig1}. Solutions for $\alpha < 0$ are not considered, since
it would result in
a complex gravitational charge $Q$, an undefined temperature $T$ as well as a negative gravitational constant.

\begin{figure}[h]
 \begin{center}
 \includegraphics[scale=0.35]{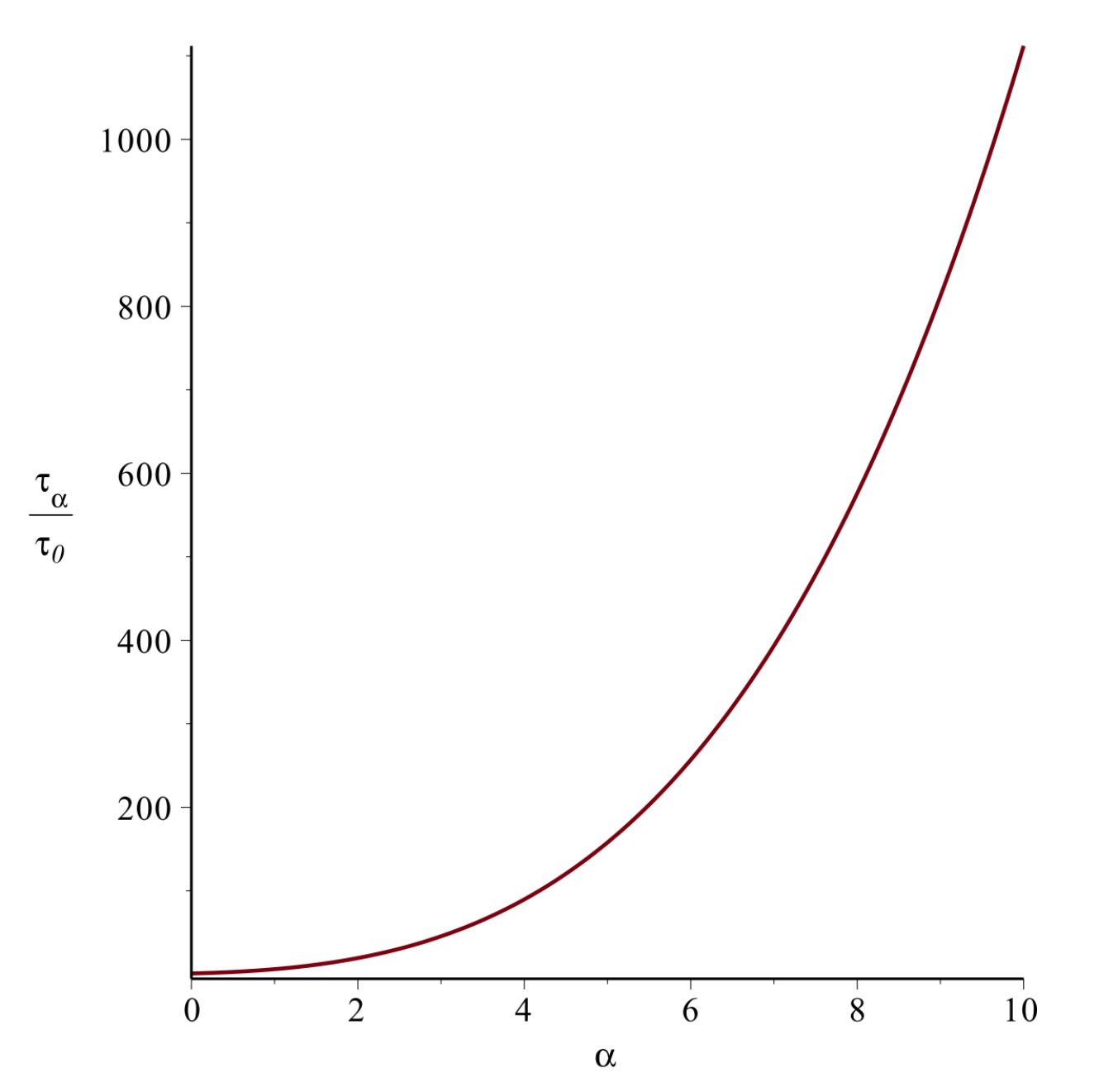}
 \caption{MOG to Schwarzschild black hole lifetime ratios $\tau/\tau_0$ for various values of $\alpha$. The gravitational constant and the 
Stefan-Boltzmann constant have been set to $G_N = \sigma= 1$.}
 \label{fig2}
 \end{center}
 \end{figure}

The radiative power of the black hole is
\beq
\frac{dM}{dt} =4\pi r_+^{2} \sigma T^4,
\eeq
where $\sigma$ is the Stefan-Boltzmann constant. The lifetime is thus
\beq
\tau= -\int_{\Mbh}^0 \frac{dM}{4\pi \sigma r_+^2 T^4}
$$ $$
= \frac{4\pi^3}{3\sigma} (1+\sqrt{1+\alpha})^4 (1+\alpha+\sqrt{1+\alpha})^2 G_N^2 \Mbh^3.
\eeq
Figure~\ref{fig2} shows the ratio of the lifetime for various values of $\alpha$ to that
of a Schwarzschild black hole ($\alpha = 0$).  The dependence on $\alpha$ clearly influences
the lifetime of such black holes.  A value of $\alpha=2$ increases the value by roughly a factor of 20.
A two-fold increase can be obtained with as small a value as $\alpha = 0.312$.

Although algebraically similar to the Reissner-Nordstr\"om black hole solution, the MOG metric
is fundamentally different in that the lack of electric charge implies that no
zero-temperature remnants may be obtained (in contrast to the extremal Reissner-Nordstr\"om case).
This is due to the fact that the gravitational charge $Q$ is proportional to $M$. We note the
``extremal limit'' $r_+ = r_-$ occurs only when $\alpha = -1$, at which point the horizons vanish.

The entropy of the black hole can be derived from the temperature by
\beq
S_{\rm T} = \int_0^{M_{\rm BH}} \frac{dM}{T}
$$ $$
= \pi G_N \Mbh^2 \left(\sqrt{1+\alpha}+1+\alpha\right)\left(\sqrt{1+\alpha}+1\right).
\label{tempentropy}
\eeq
Written in this form, it is apparent that it is related to the entropy of the Schwarzschild black hole as
given by the area-entropy law. Defining $A_+ = 4 \pi r_+^2$ and using Eq.~(\ref{moghorizons}), we find
\beq
S_{\rm A} = \frac{A_+}{4G} = \pi G_N \Mbh^2 \left(1+\sqrt{1+\alpha}\right)^2.
\label{areaentropy}
\eeq
Thus, the entropy defined in standard thermodynamics
is different from the entropy calculated using the Bekenstein-Hawking bound~\cite{1a,2}.
This can be taken as an indication that the Bekenstein-Hawking bound is modified in MOG. The MOG correction to the
Bekenstein-Hawking bound can be written as
\beq
\Delta S = S_{\rm T} - S_{\rm A}  = \pi G_N\Mbh^2 \alpha(1+\sqrt{1+\alpha}).
\eeq
The MOG leads to an interesting modification of black hole thermodynamics.
The entropy reduces to the familiar Schwarzschild black hole value $S = 4\pi G_N \Mbh^2$ when $\alpha = 0$.

The local and global thermodynamic stability of a black hole is determined by both the heat capacity and the free energy, respectively \cite{dobado1,dobado2}.
A calculation of the heat capacity for the MOG black hole yields:
\beq
C = - 2\pi G_NM^2 (1+\sqrt{1+\alpha})(1+\alpha+\sqrt{1+\alpha})  < 0~~,
\label{heatcap1}
\eeq
which typically indicates the black hole is thermodynamically unstable for all values of $M$.  The free energy $F$ can be obtained from
%in the Euclidean metric approach \cite{hawkingpage}
%as
%\beq
%F = -T\log Z
%\eeq
%where $Z \sim -\Tr e^{-M/T}$ is the partition function for the grand canonical ensemble of thermal states.  For this MOG black hole, we find...

\beq
F = M - TS 
\label{freeenergy1}
\eeq
Note the result will differ depending on the choice of the entropy expression.  Choosing the thermodynamic definition (\ref{tempentropy}), one finds
\beq
F = \frac{M}{2}
\label{mogfe1}
\eeq
while choosing the area entropy (\ref{areaentropy}) yields
\beq
F = M\left(1-\frac{1}{2}\cdot\frac{1+\sqrt{1+\alpha})}{1+\alpha+\sqrt{1+\alpha}}\right)
\label{mogfe2}
\eeq
using (\ref{tempentropy}).  Stability of the thermal system occurs when $F$ is minimized.  For (\ref{mogfe1}), this is at $M=0$, and so the system is globally unstable.  In the latter case (\ref{mogfe2}), $M=0$ is again the minimized state since $F\geq 0$ for $\alpha > 0$.  The MOG black hole is thus both locally and globally thermodynamically unstable.

\section{Thermodynamics of Rotating Black Holes in MOG}

A solution for the rotating black holes in MOG has been constructed~\cite{Moffat5}. We will call this
the Kerr-MOG solution, as it reduces to the usual Kerr solution in the limit $\alpha = 0$. The Kerr-MOG metric can be written as ($a=J/M$):
\bea
ds^2 &=& \frac{\Delta}{\rho^2} \left(dt - a\sin^2\theta\; d\phi\right)^2 - \frac{\sin^2\theta}{\rho^2} \left[(r^2+a^2)d\phi -a dt\right]^2\nonumber \\
&& -\frac{\rho^2}{\Delta} dr^2 -\rho^2 d\theta^2,
\label{kerrmog}
\eea
which describes black holes with horizon radii:
\beq
r_\pm = G_N(1+\alpha)M \left[1\pm \sqrt{1-\frac{a^2}{G_N^2 (1+\alpha)^2 M^2} - \frac{\alpha}{1+\alpha}}\right].
\label{kmhorizon}
\eeq
Moreover, $\Delta = r^2 - 2GMr+a^2+\alpha G_NGM^2$ and $\rho^2= r^2 + a^2\cos^2\theta$. The ergosphere is located at
\beq
r_e = G_N(1+\alpha)M \left[1+ \sqrt{1-\frac{a^2\cos^2\theta}{G_N^2 (1+\alpha)^2 M^2} - \frac{\alpha}{1+\alpha}}\right].
\eeq
This solution is again algebraically identical to the Kerr-Newman metric but now with the gravitational charge $Q = \sqrt{\alpha G_N}M$. The temperature for this Kerr-MOG black hole solution is determined by
\beq
T = \frac{\kappa}{2\pi}~~~,~~~\kappa = \frac{r_+ - r_-}{2(r_+^2 + a^2)}.
\eeq
We obtain
\beq
T = \frac{1}{2\pi G_NM} \cdot \frac{\beta}{2\beta (\alpha+1) +\alpha+2},
\label{kmtemp}
\eeq
where
\beq
\beta = \sqrt{\frac{\alpha + 1 - \frac{a^2}{G_N^2 M^2}}{(1+\alpha)^2}}
\eeq
In contrast to the spherically symmetric static MOG black hole, the Kerr-MOG back holes possesses an extremal configuration when $r_+ = r_-$, or
\beq
1-\frac{a^2}{G_N^2 M^2 (1+\alpha)^2} - \frac{\alpha}{1+\alpha}=0
$$ $$
~~~\Longrightarrow~~~M_{\rm ext} = \frac{1}{G_N} \frac{a}{\sqrt{1+\alpha}}.
\label{extmass}
\eeq

 \begin{figure}[h]
 \begin{center}
 \includegraphics[scale=0.35]{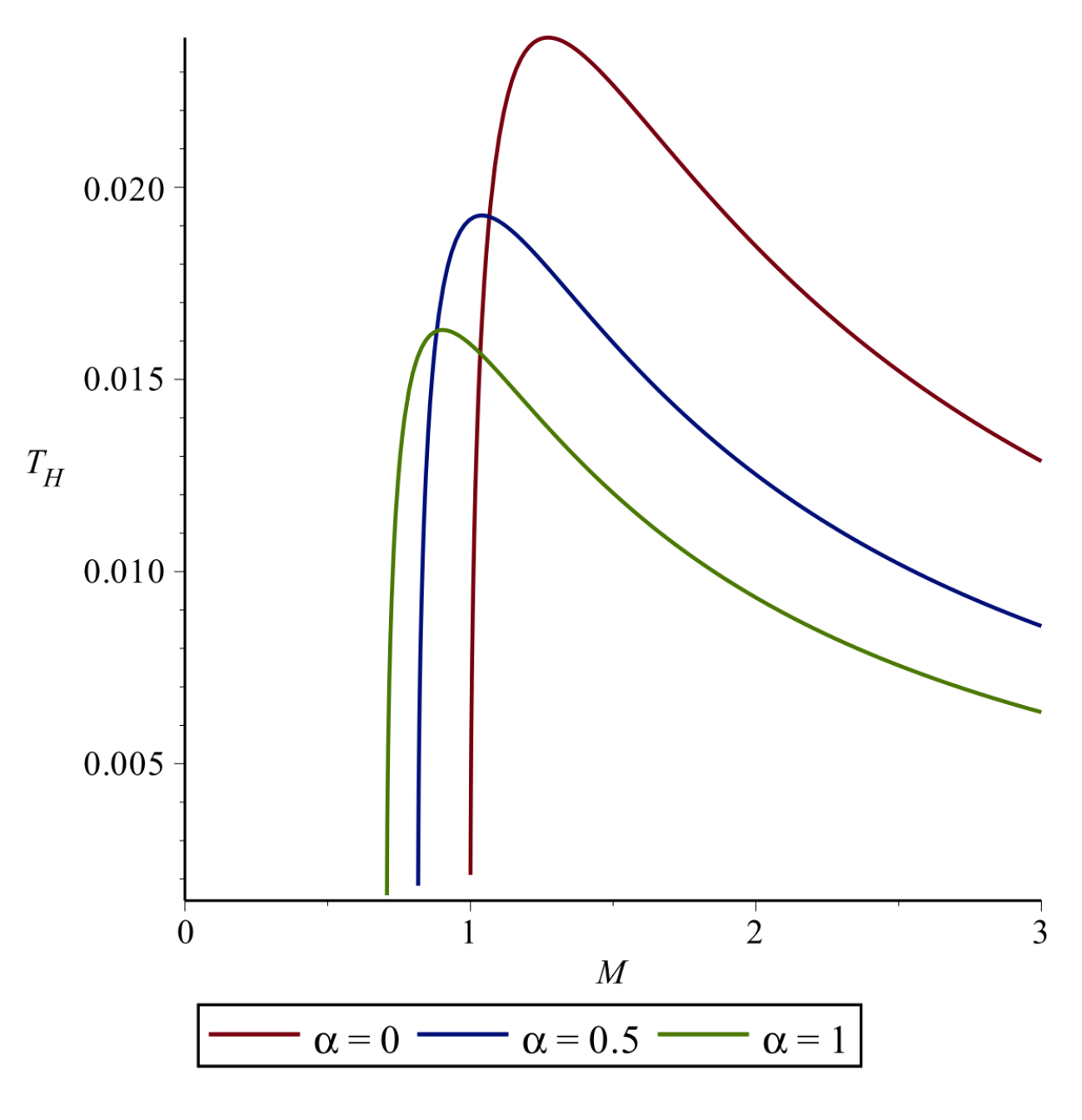}
 \caption{Kerr-MOG temperature profile for spin parameter $a=1$ and $\alpha = 0, 0.5, 1$.  The remnant mass values for each curve are $M_0 = 1, 0.816, 0.707$ for $\alpha = 0, 0.5, 1$, respectively.  These reach maximum temperatures of  $T_{\rm max} = 0.023, 0.019, 0.016$ for $M_{\rm max} = 1.272, 1.040, 0.902$, respectively.  The curves do not appear to end at $T_H = 0$ due to plotting resolution.}
 \label{kerrmogtemps}
 \end{center}
 \end{figure}

 \begin{figure}[h]
 \begin{center}
 \includegraphics[scale=0.35]{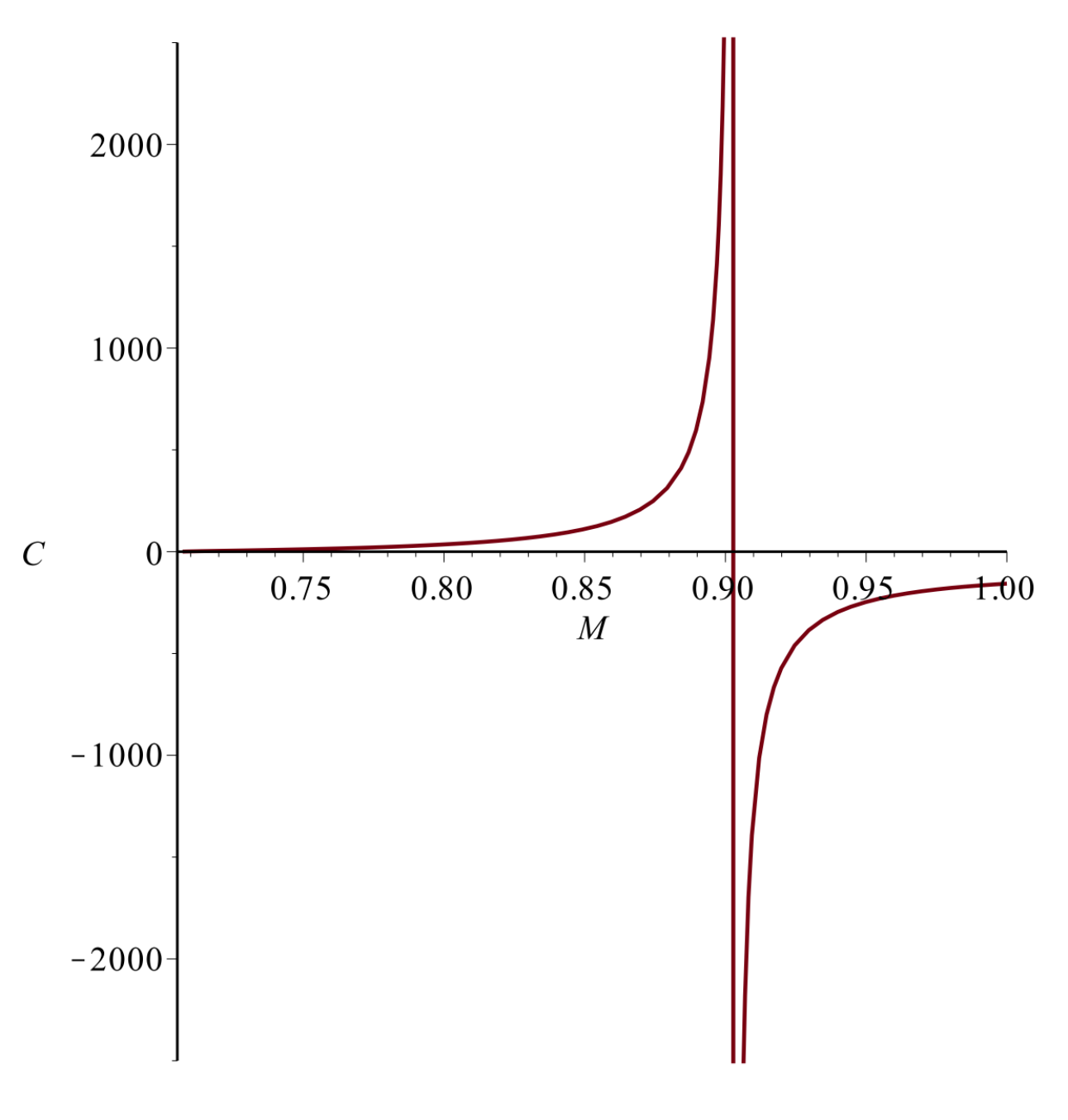}
 \caption{Kerr-MOG heat capacity for $\alpha = 1$ and $a=1$.  The black hole is thermodynamically unstable when $C<0$, and reaches a maximum temperature when $C\rightarrow \pm \infty$ ($M_{\rm max} \approx 0.9$ in this case).  For $M<M_{\rm max}$, the heat capacity is positive and the black hole evaporates to a locally stable cold remnant.}
 \label{kmhc}
 \end{center}
 \end{figure}

\begin{figure}[h]
 \begin{center}
 \includegraphics[scale=0.35]{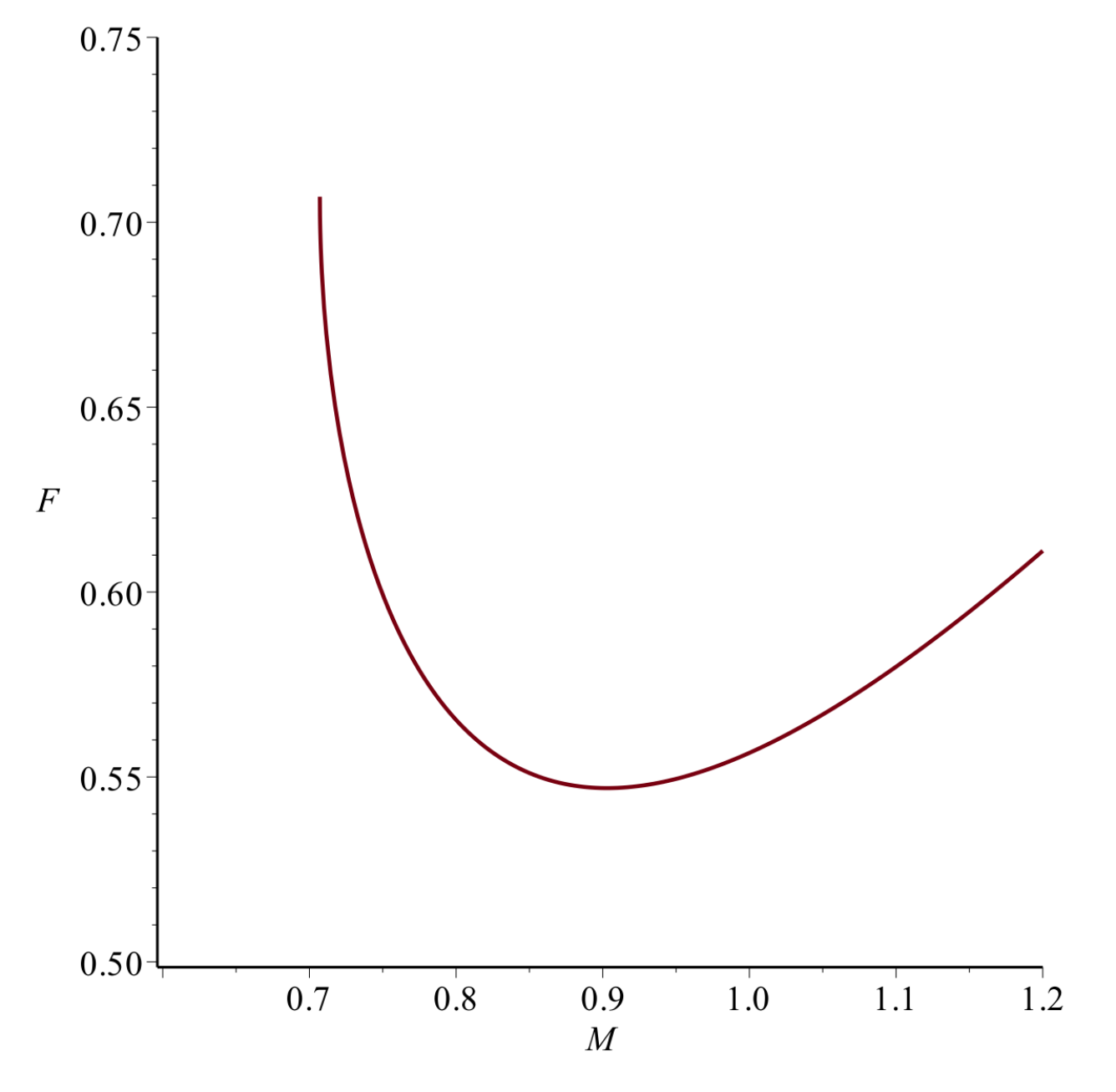}
 \caption{Kerr-MOG Gibbs free energy for $\alpha = 1$ and $a=1$.  The curve reaches a minimum near $M\sim 0.9$, consistent with the transition point of the heat capacity. }
 \label{kmfree}
 \end{center}
 \end{figure}

The Kerr-MOG black hole free energy can be calculated from (\ref{freeenergy1}), using the temperature (\ref{kmtemp}) and entropy defined by (\ref{tempentropy}), and is
displayed in Figure~\ref{kmfree}.  The free energy is minimized at the same point that defines the transition point for the heat capacity.  This strengthens the assertion that this represents a critical point in the thermodynamic stability, and the positivity of the free energy indicates global instability so the black hole undergoes complete evaporation.

It can be shown that the temperature (\ref{kmtemp}) vanishes for
a value of $\alpha$, implying that this is a zero-temperature remnant.
Corresponding temperature profiles are plotted in Figure~\ref{kerrmogtemps}, and the generic behavior
of the heat capacity is shown in Figure~\ref{kmhc}. The effect for increasing $\alpha> 0$ is to shift the temperature curve to the left, resulting in small remnants. Furthermore, the maximum temperature of the Kerr-MOG black hole is correspondingly decreased.

\section{Regular MOG Black Hole}

In this section, we will analyze the thermodynamics of regular
solutions which occur in the full non-linear MOG action. It may be noted that
a regular solution for electrically charged objects in GR has been constructed using non-linear
electrodynamics~\cite{Garcia}. As the kinetic part of the action of the vector field in MOG
can be taken to be non-linear, it is possible to  construct
  a regular black hole solution in MOG~\cite{Moffat5}.
This MOG regular solution can have two horizons, an outer horizon $r_+$
and an inner Cauchy horizon $r_-$ for $\alpha < \alpha_{crit} = 0.673$. However, for $\alpha >
\alpha_{crit}$ there are no horizons and no naked singularity exists. The case without a horizon is called a
 ``gray hole'', and the gravity of this gray hole for a sufficiently large mass $M$
is expected to be so strong that it effectively can be treated
as a black hole. The redshift observed by an asymptotic external observer
will be large and there is small probability for any object to leave its gravitational field.
However, as this probability is not zero, it will not suffer from the black hole information loss
paradox~\cite{Hawking2,Giddings1,Giddings2}. So, it is interesting to note that the modification of gravity
which was phenomenologically proposed to explain the correct dynamics of galaxies might end up predicting
the absence of a horizons for objects in our universe, and hence end up solving the black hole
information loss paradox for these astrophysical objects.

\begin{figure}[h]
\begin{center}
\includegraphics[scale=0.35]{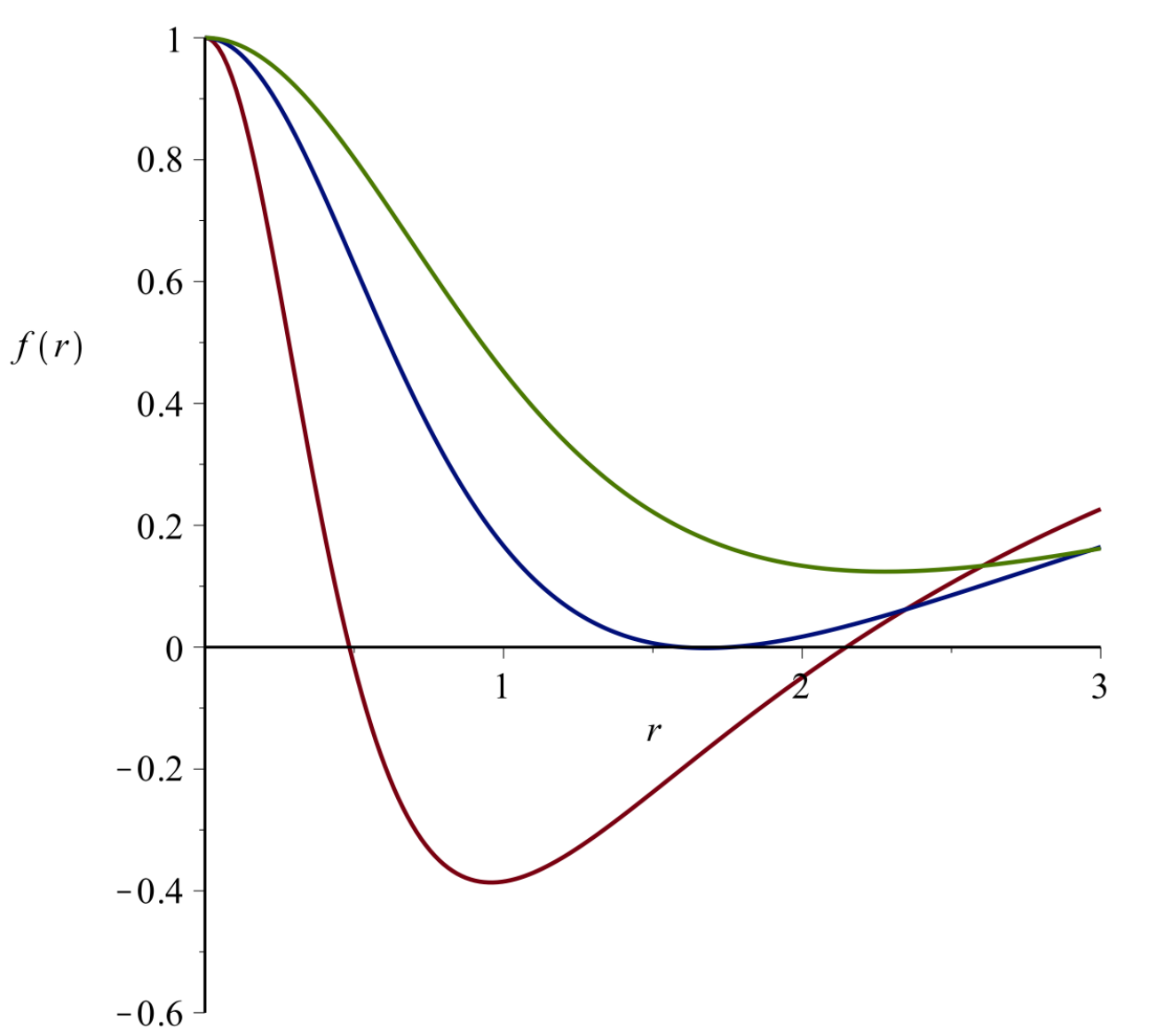}
\caption{Regular MOG metric function for values $\alpha = 0.3$ (lower curve), $\alpha = \alpha_{\rm crit} \approx 0.673$ (middle curve), and $\alpha =1$ (upper curve). The black hole mass is normalized to $M = 1$, as is the gravitational constant $G_N = 1$. The radial value for which $f(r_{\rm crit}) = 0$ is $r_{\rm crit} \approx 1.68$.  Note that the value of $\alpha_{\rm crit}$ is independent of the mass.}
\label{fig3a}
\end{center}
\end{figure}

The metric function for the regular solution can be written as~\cite{Moffat5}:
\beq
ds^2 =  f(r) dt^2 - \frac{dr^2}{f(r)} - r^2 d\Omega^2,
\eeq
where
\beq
f(r) = 1-\frac{2GMr^2}{\left(r^2+\alpha G_N GM^2\right)^{3/2}}+\frac{\alpha GG_N M^2r^2}{(r^2+\alpha GG_NM^2)^2}.
\label{regularmog}
\eeq
It describes a Schwarzschild-MOG black hole for large $r$ and the metric becomes asymptotically
flat in the limit $r\rightarrow \infty$.  For small $r$, it describes a de Sitter (anti-de Sitter) core
depending on the value of $\alpha$ with the effective cosmological constant:
\beq
\Lambda = \frac{3}{G_N^2 M^2}\left(\frac{\alpha^{1/2} -2}{\alpha^{3/2}(1+\alpha)}\right).
\eeq
The metric function $f(r)$ approaches $1$ in the limit $r=0$.  Horizons are obtained as the roots of $f(r)=0$, which depend on the 
critical value $\alpha=\alpha_{\rm crit}=0.673$. Figure~\ref{fig3a} shows the behavior of $f(r)$ for a normalized mass $M = 1$ and $G_N=1$, 
in which the three phases of solutions are visible.

A gray hole can approach a black hole with horizons as a limiting process, when $\alpha \to \alpha_{crit}$.
In general, the thermodynamic properties of the gray hole cannot be the same as a MOG black hole possessing horizons. It is expected that 
a gray hole's entropy will be less than the entropy of a black hole of a similar size and mass, for it is not a maximum entropy object. 
It is possible to study the thermodynamic features of the regular solution as $\alpha\rightarrow\alpha_{\rm crit}$, 
when horizons form and an asymptotic observer will see that the redshift of light will approach infinity.

If MOG is the correct theory describing gravity, then it is possible that only a regular solution without horizons forms during gravitational collapse. 
Because there are no horizons the gray hole will have negligible or no Hawking radiation and there is no information loss paradox. 
However, we must caution that at present, we have no known physical law that demands that $\alpha > \alpha_{\rm crit}$.

In order to evaluate the temperature for the regular black hole with $\alpha < \alpha_{\rm crit}$, we proceed by the surface gravity method.  The derivative of the metric function is given by
\bea
 f^\prime(r)& =& -\frac{4(1+\alpha)Mr}{\zeta^{3/2}}+\frac{6(1+\alpha)M r^3}{\zeta^{5/2}}\nonumber \\
 &&+\frac{2\alpha(1+\alpha)M^2r}{\zeta^2}-\frac{4\alpha(1+\alpha)M^2 r^3}{\zeta^3},
\eea
where $\zeta=r^2+\alpha(1+\alpha)M^2$ and the derivative is evaluated at the outer horizon $r = r_+$.  The corresponding real, positive roots of $f(r) =0$ are given in Table~\ref{table2}. Evaluating the temperature $T = f^\prime(r=r_+)/4\pi$ for the values therein, one observes that $T \sim 1/M$ for all values of $r$.  This means that in the presence of horizons, the temperature diverges as the black hole evaporates, despite the regular nature of the solution at $r=0$. Thus, the divergent behavior of the temperature $T$ in the singular Schwarzschild black hole solution is retained for the regular black hole solution with an outer horizon.

\begin{table}[h]
\begin{center}
\begin{tabular}{ccc}
$\alpha$ & $r_-$ & $r_+$ \\ \hline
$0.1$ & $0.163 M$ & $2.067 M$ \\
$0.2$ &$0.135 M$ & $2.117 M$ \\
$0.3$ & $0.438 M$ & $2.148 M$ \\
$0.4$ & $0.676 M$ & $2.152 M$ \\
$0.5$ & $0.904 M$ & $2.120 M$ \\ \hline
\end{tabular}
\caption{Inner and outer horizons $r_\pm$ for the regular MOG black hole for various values of $\alpha >0$. }
\label{table2}
\end{center}
\end{table}

\section{MOG Black Hole Entropy Corrections}

It is expected that quantum fluctuations in the geometry will lead to thermal
fluctuations in the thermodynamics of black holes. The explicit form of such corrections can be calculated by analyzing
the thermal fluctuations around a state of equilibrium~\cite{l1, SPR}. This can be done using the path integral formalism.
In the path integral formalism, it is possible to calculate the amplitude for a field configuration to propagate to another field configuration. 
This can be done using the Euclidean quantum gravity formalism, where the temporal coordinate is rotated in  the complex plane
~\cite{o, 01ab, 02ab, 04ab, 05ab}. Thus, the partition function for spacetime can be written as
\begin{equation}
Z = \int [D]   \exp( - S_E),
\end{equation}
where $S_E = -i S$ is the Euclidean action corresponding to the MOG action $S$. 
Thus, $S_E$ is obtained from $S$ by a rotation of the time axis in the complex plane.

It may be noted that the partition function can be related to the statistical mechanical partition function~\cite{hawk,hawk1}:
\begin{equation}
Z = \int_0^\infty  dE\rho (E)\exp(-\beta E),
\end{equation}
where $\beta$ is the inverse of the temperature. The density of states can be written as
\begin{eqnarray}
\rho (E) = \frac{1}{2 \pi i} \int^{\beta_0+ i\infty}_{\beta_0 - i\infty}d \beta\exp[S(\beta)],
\end{eqnarray}
where
\begin{equation}
S = \beta  E + \ln Z.
\end{equation}
Usually this entropy is measured around the equilibrium temperature $\beta_0$, and all thermal fluctuations are
neglected. This is done by making the identification $T = \beta^{-1}$, and $S_0 = S(\beta)_{\beta = \beta_0} = A/ 4G_N$, 
where as before $A$ is the area of the black hole~\cite{th}. 
Identifying the temperature in (\ref{mogtemp1}) with $\beta_0$ and using the entropy expressions (\ref{tempentropy})
and (\ref{areaentropy}) for $\tilde{S}(\beta_0)$ and $S(\beta_0)$, respectively, we can write the difference:
 \begin{eqnarray}
\Delta S (\beta_0) &=&  \tilde S (\beta_0)- S(\beta_0) \nonumber\\
&=& \pi G_N M^2 \alpha (1 + \sqrt{ 1 + \alpha}).
\end{eqnarray}

Taking such thermal fluctuations into account, we can
expand $S(\beta)$ around the equilibrium temperature $\beta_0$~\cite{l1, SPR} to obtain
\begin{equation}
S = S_0 + \frac{1}{2}(\beta - \beta_0)^2
\left(\frac{\partial^2 S(\beta)}{\partial \beta^2 }\right)_{\beta = \beta_0}, 
\label{a1}
\end{equation}
where we have neglected higher order corrections to the entropy and defined $S_0 = S(\beta)|_{\beta = \beta_0}$. The density of states can now be written as
\begin{eqnarray}
\rho (E) &=& \frac{\exp(S_0)}{2 \pi i}\int^{\beta_0+ i\infty}_{\beta_0 - i\infty}d\beta \nonumber \\  &&
\times
\exp \left( \frac{1}{2}(\beta-\beta_0)^2 \left(\frac{\partial^2 S(\beta)}{\partial \beta^2 }
\right)_{\beta = \beta_0}   \right).
\end{eqnarray}
After a change of variables, we get
\begin{equation}
\rho(E) = \frac{\exp(S_{0})}{\sqrt{2\pi}} \left[\left(\frac{\partial^2
S(\beta)}{\partial \beta^2 }\right)_{\beta = \beta_0}\right]^{- 1/2},
\end{equation}
and it follows that
\begin{equation}
S = S_0 -\frac{1}{2}
\ln \left[\left(\frac{\partial^2 S(\beta)}{\partial \beta^2 }\right)_{\beta= \beta_0}\right].
\end{equation}
The second derivative of entropy is actually a fluctuation squared of the energy.

It is possible to simplify this expression by using the microscopic degrees of freedom calculated
from a conformal field theory. The modular invariance of the conformal field's partition
function constraints take the form $ S(\beta) = a \beta   + b \beta^{-1}$~\cite{card}, which can be generalized to the expression: 
$S(\beta) = a \beta^m   + b \beta^{-n}$~\cite{l1, SPR},
where all the constants satisfy, $m, n, a, b > 0$. This has an extremum at $\beta_0 = (nb/ma)^{1/ m+n} = T^{-1}$.

Expanding the entropy around this extremum, we get
\begin{eqnarray}
S(\beta) &=& [(n/m)^{m/(m+n)} + (m/n)^{n/(m+n)} ](a^n b^m)^{1/(m+n)}
\nonumber \\
&&
+ \frac{1}{2}[(m+n) m^{(n+2)/(m+n)} n^{(m-2)/(m+n)}]
\nonumber \\ && \times
( a^{n+2}b^{m-2})^{{1}/(m+n)}(\beta - \beta_0)^2. \label{a2}
\end{eqnarray}
By comparing Eq. (\ref{a1}) to Eq. (\ref{a2}), we obtain
\begin{equation}
S_0 = (n/m)^{m/(m+n)} + (m/n)^{n/(m+n)} (a^n b^m)^{{1}/(m+n)},
\end{equation}
and thus
\begin{eqnarray}
\left(\frac{\partial^2 S(\beta)}{\partial \beta^2 }\right)_{\beta = \beta_0} &=& (m+n) m^{(n+2)/(m+n)}
n^{(m-2)/(m+n)} \nonumber \\ && \times ( a^{n+2}b^{m-2}  )^{1/(m+n)}~~.
\end{eqnarray}
We can now find the values of $a, b$, and simplify the expression to obtain:
\begin{eqnarray}
\left(\frac{\partial^2 S(\beta)}{\partial \beta^2 }\right)_{\beta = \beta_0} =  \mathcal{Y}_{mn} S_0 T^2~,
\end{eqnarray}
where
\begin{eqnarray}
\mathcal{Y}_{mn} &=& \left[  \left(\frac{(m+n)
m^{(n+2)/(m+n)} n^{(m-2)/(m+n)}}{(n/m)^{m/(m+n)} + (m/n)^{n/(m+n)} }
\right)~\right. \nonumber \\  && \left. \times
\left(\frac{n}{m} \right)^{2/(m+n)}\right].
\end{eqnarray}
As the factors $\mathcal{Y}_{mn}$ are independent of any black hole
parameters, they can be absorbed by a suitable redefinition (see~\cite{l1, SPR}).
We will use the corrected form for the entropy, which neglecting higher order corrections can be written as
\begin{equation}
S=S_{0}-\frac{1}{2}\ln{S_{0}T^{2}}.
\end{equation}
We can now write the the corrected entropy as
\begin{eqnarray}
S_{\rm A} &=& \pi G_N  M^2  ( 1 + \sqrt{1 + \alpha}  )^2
\nonumber \\ &&- \frac{1}{2} \ln
\frac{1}{4\pi G_N }  \frac{1}{ (1+\alpha+\sqrt{1+\alpha})^2 }.
\end{eqnarray}
This is the expression for the corrected entropy of a static black hole in MOG, due to thermal fluctuations.
If we use the definition of the entropy from its temperature, we obtain
\begin{eqnarray}
\tilde S_{\rm T} &=& \pi G_N  M^2 (\sqrt{\alpha +1 } + 1 + \alpha )(\sqrt{\alpha + 1} +1 )
\nonumber \\ && -
\frac{1}{2}\ln\biggl[\frac{1}{(\sqrt{\alpha +1 } + 1 + \alpha )(\sqrt{\alpha + 1} +1 )}\biggr].
\end{eqnarray}
We also have
\begin{eqnarray}
\Delta S &=&   S_{T 0} - S_{A 0}  - \frac{1}{2} \ln\biggl(\frac{S_{ T 0} }{S_{A 0}}\biggr)\nonumber \\
&=& \pi G_N M^2 \alpha (1 + \sqrt{ 1 + \alpha})\nonumber \\
&& - \frac{1}{2}\ln\biggl[\frac{(1 + \alpha + \sqrt{1+\alpha})}{(1+\sqrt{1+\alpha})}\biggr].
\end{eqnarray}

The corrections are the standard logarithmic corrections for the entropy of the MOG black hole.
They reduce to the entropy of a black hole expected from the holographic principle. 
Hence, these thermal fluctuations also lead to a violation of the holographic principle 
for these black holes. This is because the entropy of the black hole is reduced due to thermal 
fluctuations. This term by which the original entropy of the black hole is reduced is proportional to 
$\ln  S_0 T^2$. It may be noted that as there are two different ways of defining 
MOG entropy i.e., $S_{T 0}$ and $S_{A 0}$, the exact value of this term depends on such a definition chosen. 
However, in both these cases, the entropy of the MOG black hole reduces due to thermal fluctuations, and so these 
black holes will have less entropy than what would be expected from the holographic principle. Such violation of the 
holographic principle only occurs at small scale, where the temperature of the MOG black hole is sufficiently large, and 
so the effect of thermal fluctuations cannot be neglected. It may be noted that at this stage, the effects of 
quantum fluctuations can also not be neglected. In fact, the quantum fluctuations in the geometry of the black hole lead 
to these thermal fluctuations in the thermodynamics of the black hole. 
The violation of the holographic principle due to quantum fluctuations has been 
studied previously~\cite{6, 6a}. We have demonstrated explicitly that such an effect occurs for MOG black holes due to thermal fluctuations.

\section{Conclusions}

We have analyzed the thermodynamics of black holes in a modified theory of gravity that contains
a repulsive vector field $\phi_\mu$ with the gravitational charge
$Q=\sqrt{\alpha G_N}M$. In the absence of matter, the field equations have a non-zero
energy-momentum tensor formed from the $B_{\mu\nu}$ field. For non-vanishing values of the
parameter $\alpha$ there are no pure vacuum solutions of the modified gravitational field equations.
The thermodynamic properties of the solutions corresponding to a spherically symmetric black hole and
a rotating black hole were analyzed, and it was found that the the entropy area law gets changed by
increasing the size of the MOG parameter $\alpha$. The thermodynamics of a regular black hole solution
in MOG with $\alpha < \alpha_{crit}$ with two horizons was investigated. We also considered a regular
solution without horizons describing a ''gray`` hole. There is a small but finite probability for 
particles to escape the gray holes. Thus, the modification of gravity which was initially 
proposed to explain the dynamics of galaxies and galaxy clusters without dark matter could 
end up predicting the absence of horizons for massive collapsed objects in our universe. 

We also analyzed the corrections to the thermodynamics of a black hole in MOG.
The standard partition function for a static black hole in the modified theory of gravity was 
used to perform this analysis. Thus, the thermodynamic properties of a MOG black hole were obtained using this partition
function, and the thermal fluctuations to the thermodynamic quantities were computed.
Motivated by a conformal field theory description, an explicit form for the corrections to the entropy of
the static black hole was found. It was demonstrated that corrections had the standard logarithmic form.
The entropy of the black hole is less than what is expected from the holographic principle. Hence, 
the thermal fluctuations also lead to a violation of the holographic principle for static black holes in MOG.

%Note that in the Jacobson approach the Einstein field equations are derived from
%the Clausius relation~\cite{jaco}, and a modification to the entropy will modify the Einstein field equations
%and solutions to these equations. 
It has been observed that a traversable MOG wormhole exists as a solution of the MOG field equations~\cite{Moffat5}. So, it will be interesting to analyze the effects of the corrections to the MOG thermodynamics for such a wormhole solution.

Corrections to ordinary GR black holes have been obtained using the generalized uncertainty principle~\cite{mi}.
It was demonstrated that such corrections can lead to the existence of black hole remnants, which can have important phenomenological consequences~\cite{r1}. It will be interesting to analyze such corrections for MOG black holes. Additional interesting and testable phenomenological characteristics can be extracted from the
MOG black hole solutions, including quasinormal modes and gravitational wave signatures of binary inspirals. The latter is of particular interest due to the definitive fingerprints of MOG in a black hole's shadow~\cite{Moffat6}, which are expected to be detectable by the Event Horizon Telescope observations. If similar MOG effects are as pronounced for gravitational waves, then they should be detectable in the upcoming run of LIGO~\cite{aligo}.   Lastly, we can also consider other types of non-linear kinetic 
terms for MOG ~\cite{Garcia, non, non1, non2, non4,
non5, non6}
\begin{eqnarray}
\mathcal{K}_1&=& (- \mathcal{B})^s, \nonumber \\
\mathcal{K}_2&=& 4 b^2  (1 - \sqrt{1 - (2b)^{-1} \mathcal{B} }), \nonumber \\
\mathcal{K}_3 &=& b^2 (\exp (-b^{-2} \mathcal{B}) - 1), \nonumber \\
\mathcal{K}_4 &=& - 8 b^2 \ln (1 + b^{-2} \mathcal{B}),
\end{eqnarray}
where $s, b$ are two parameters. It would be interesting to analyze the phenomenological consequences of such 
terms.  %Such analyzes are currently underway by the authors.

\vskip 1cm
\noindent {\bf Acknowledgements}\\

JRM thanks the generous hospitality of the Perimeter Institute for Theoretical Physics, where this work was commenced. Research at the Perimeter Institute for Theoretical Physics is supported by the Government of Canada through industry Canada and by the Province of Ontario through the Ministry of Research and Innovation (MRI).

\end{document}